\documentclass{article}
\usepackage[pdftex]{graphicx}

\usepackage[T1]{fontenc} 
\usepackage[utf8]{inputenc} 
\usepackage{ismir,amsmath,cite,url}
\usepackage{graphicx, booktabs, multirow}
\usepackage{color}
\usepackage{booktabs} 
\usepackage{caption}  
\usepackage{float}
\usepackage{units}
\usepackage{paralist}
\usepackage{nicefrac}
\usepackage{microtype}

\usepackage{tabularx} 
\usepackage{adjustbox} 
\usepackage{booktabs}
\usepackage[firstpage]{draftwatermark}
\definecolor{lightgray}{rgb}{0.9,0.9,0.9}
\definecolor{darkgray}{rgb}{0.4,0.4,0.4}
\SetWatermarkFontSize{12pt}
\SetWatermarkScale{1.1}
\SetWatermarkAngle{90}
\SetWatermarkHorCenter{202mm}
\SetWatermarkVerCenter{170mm}
\SetWatermarkColor{darkgray}
\SetWatermarkText{Late-Breaking / Demo Session Extended Abstract, ISMIR 2024 Conference}

\def\lbd{}	
\title{Towards Robust Transcription: Exploring Noise Injection Strategies for Training Data Augmentation}

\oneauthor
{Yonghyun Kim \hspace{1cm} Alexander Lerch}
{Music Informatics Group, Georgia Institute of Technology, USA\\ {\tt\small \{yonghyun.kim, alexander.lerch\}@gatech.edu} }

\def\authorname{Y. Kim, and A. Lerch}

\begin{document}

\maketitle
\begin{abstract}
Recent advancements in Automatic Piano Transcription (APT) have significantly improved system performance, but the impact of noisy environments on the system performance remains largely unexplored.
This study investigates the impact of white noise at various Signal-to-Noise Ratio (SNR) levels on state-of-the-art APT models and evaluates the performance of the Onsets and Frames model when trained on noise-augmented data. We hope this research provides valuable insights as preliminary work toward developing transcription models that maintain consistent performance across a range of acoustic conditions.

\end{abstract}
\section{Introduction}\label{sec:introduction}

Automatic Music Transcription (AMT) is a fundamental task in Music Information Retrieval (MIR), aiming to convert audio recordings into symbolic representations, such as the Musical Instrument Digital Interface (MIDI) format. 
As a foundational task in Music Information Retrieval (MIR), AMT has broad applications in music education, search, creation, and musicology \cite{Benetos2019}.
While AMT in general targets all musical instruments and efforts have been made for a variety of instruments \cite{Wu2018, Wu2020, ICLR22Gardner:01}, the majority of research has focused on transcribing solo piano performances \cite{ISMIR18Hawthorne:01, ISMIR20Kwon:01, ICLR19Hawthorne:01, Kong2021, NIPS21Yan:01, ISMIR22Wei:01, ISMIR23Toyama:01, Kwon2024}. Recent advancements in piano transcription have led to significant improvements, particularly in onset detection; models trained on the MAESTRO dataset \cite{ICLR19Hawthorne:01} have achieved onset F1 scores exceeding 95\% \cite{Kong2021, NIPS21Yan:01, ICLR22Gardner:01, ISMIR22Wei:01, ISMIR23Toyama:01, Kwon2024}. However, these results are typically based on evaluations using very clean data. Thus, the robustness of these piano transcription systems in real-world scenarios remains largely unexplored. Some previous work shows how performance degrades in altered acoustic environments \cite{Edwards2024} and how robustness can be improved through data augmentation \cite{ICLR19Hawthorne:01, Edwards2024}.

Despite the proven benefits of data augmentation in improving model robustness, there are no established guidelines for prioritizing specific techniques in AMT. While various studies have explored different data augmentation methods \cite{ICLR19Hawthorne:01, Edwards2024, ICASSP18Thickstun:01}, there is limited clarity on how to apply these techniques effectively. For instance, when employing noise injection, several key factors must be considered, such as the type of noise (e.g., white, pink, environmental), the Signal-to-Noise-Ratio (SNR), and the ratio of clean to augmented data. However, to the best of our knowledge, these parameters are often chosen arbitrarily, highlighting the need for further investigation in this area.

In this study, we investigate the impact of white noise injection, a widely adopted technique in audio research \cite{NIPS21Lim:01, ICASSP22Conti:01, ICLR23Eberhard:01}, on piano transcription. Our analysis has two primary goals,
\begin{inparaenum}[(i)]
    \item to assess the performance degradation of AMT systems across various SNR levels, and 
    \item to demonstrate the effectiveness of white noise injection in enhancing the robustness of the transcription.
\end{inparaenum}

\section{Experiments}
This section outlines the experimental methods and summarizes the results. Due to the page limit constraints, detailed numerical results are not included in the main text but can be accessed, along with the experimental code, on GitHub.\footnote{https://github.com/yonghyunk1m/TowardsRobustTranscription}

\subsection{Exp.~1: Impact of Noise on Pre-Trained Systems}\label{sec:exp1}

Recording piano performances in uncontrolled environments often results in widely varying SNRs due to lack of high quality equipment, background noise, reverberation, and other environmental factors. In such settings, the SNR tends to be lower than in controlled professional environments such as studios. To rigorously assess the robustness of the models under these conditions, we evaluated their performance across a broad range of SNR values, ranging from \unit[-6]{dB} to \unit[45]{dB}, in \unit[3]{dB} intervals, covering a total of 18 different SNR levels.

For this experiment, we utilized two state-of-the-art models: Onsets and Frames \cite{ISMIR18Hawthorne:01} and the model proposed by Kong et al.~\cite{Kong2021}. The Kong et al.\ model was tested using a publicly available pre-trained checkpoint,\footnote{https://github.com/bytedance/piano\_transcription} while the Onsets and Frames \cite{ISMIR18Hawthorne:01} model was re-implemented using the PyTorch framework,\footnote{https://github.com/jongwook/onsets-and-frames} following the original training procedures and specifications outlined in the respective papers.

We generated white noise-augmented versions of the MAESTRO v3 \cite{ICLR19Hawthorne:01} dataset for inference. The test split, consisting of 177 recordings, was used for evaluation as defined in the dataset's metadata. To create the augmented data, we calculated the power of each target audio sample and added scaled white noise to achieve the desired SNR levels. The resulting audio was then normalized to retain the RMS value of the original, noise-free recording. In instances where clipping occurred, we adjusted the affected sections to the maximum representable value to prevent distortion, ensuring that the integrity of the audio signal was preserved throughout the evaluation process.

\begin{figure} [t]
 \centerline{
\includegraphics[width=0.459\textwidth]{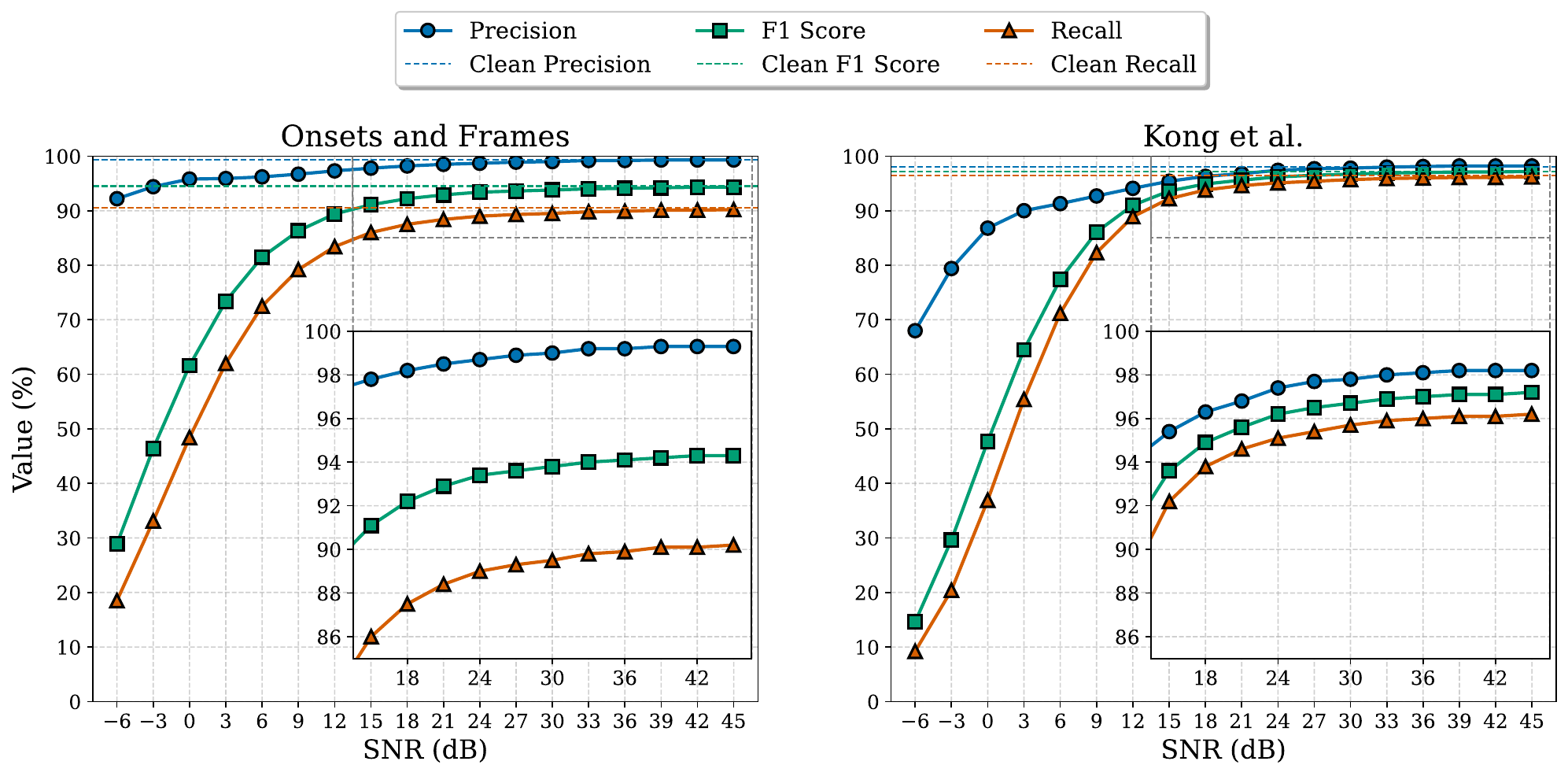}}
 \caption{Inference results of the Onsets and Frames model and the Kong et al.\ model, evaluated on the white noise-injected MAESTRO test split across varying SNR levels.}
 \label{fig:1}
\end{figure}

\figref{fig:1} presents the overall results of this experiment. As the SNR level decreases, a clear decline in model performance is observed. Both the Kong et al. \cite{Kong2021} model and Onsets and Frames \cite{ISMIR18Hawthorne:01} show a reduction in F1 scores compared to their baseline performance on clean data (96.7\% and 94.5\%, respectively), with around a 5\% relative drop at \unit[12]{dB} SNR and around a 10\% relative drop at \unit[9]{dB} SNR.

\subsection{Exp.~2: Effect of Noise Injection during Training}

To investigate whether performance degradation in noisy environments can be mitigated through data augmentation, we employed the Onsets and Frames model \cite{ISMIR18Hawthorne:01}, training it with white noise-injected audio. We introduce the term Clean-to-Noise Ratio (CNR) to represent the proportion of clean to noise-injected audio sampled during training. Based on the hypothesis that the level of perturbation impacts performance, we trained the model using CNR levels of 0 (fully perturbed), 1/3, 1, 3, and $\infty$ (clean audio only). For each CNR, the model probabilistically samples from the clean or noisy audio. In case of noisy audio, the SNR dB value is randomly sampled from [0, 24], followed by RMS normalization and clipping prevention. This SNR range was chosen based on prior research \cite{ICLR19Hawthorne:01, Edwards2024} and results from Sect.~\ref{sec:exp1}, where significant performance degradation was observed.

We followed the original Onsets and Frames \cite{ISMIR18Hawthorne:01} configuration throughout the training process, standardizing the number of iterations to 100k for all experiments. This decision was based on preliminary experiments with noise-injected audio, which showed that increasing the number of iterations beyond 100k did not result in lower losses. 

\begin{figure}
 \centerline{
 \includegraphics[width=0.5\textwidth]{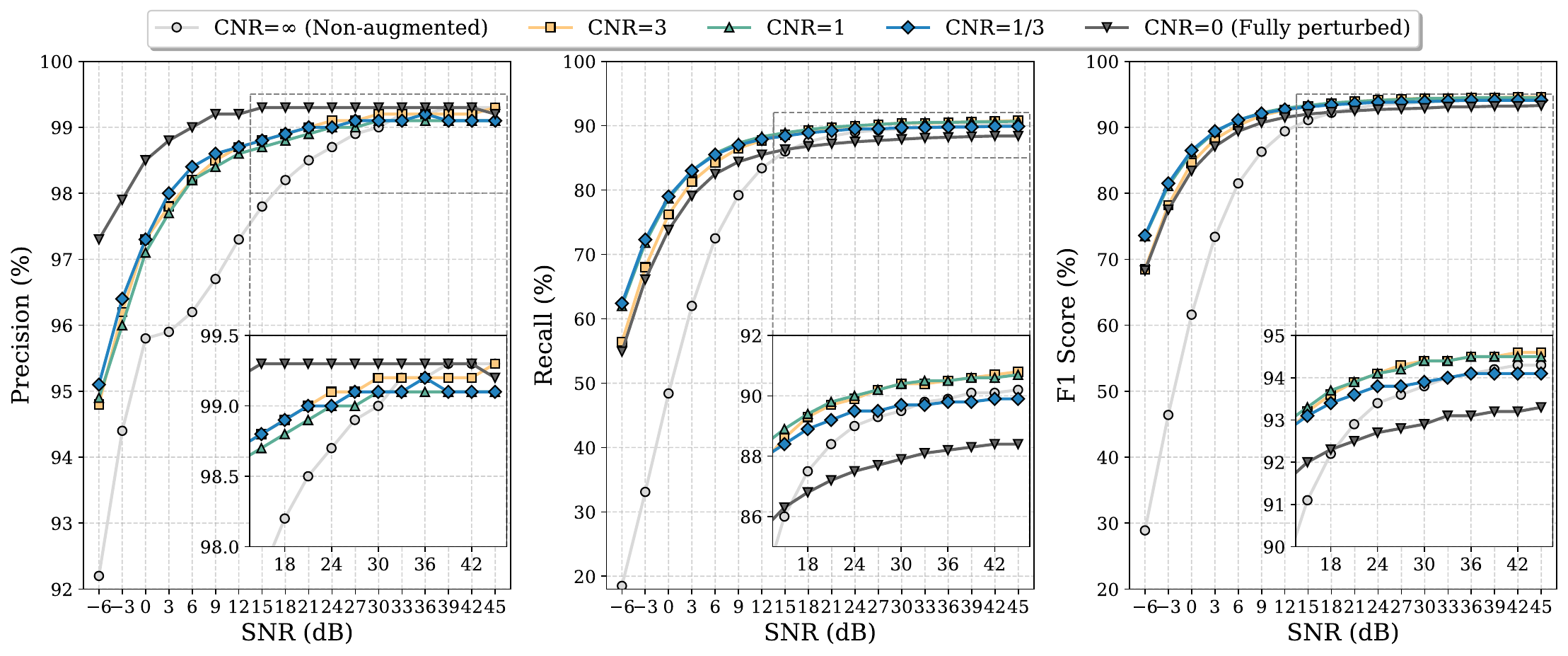}}
 \caption{Inference results for the Onsets and Frames model across different CNR values, evaluated on the white noise-injected MAESTRO test split at varying SNR levels.}
 \label{fig:2}
\end{figure}

The performance evaluation of the trained models is presented in \figref{fig:2}.

We observe that the the systems trained with noisy data all considerably outperform the system trained with clean data for low SNRs. As the SNR increases, the gap decreases, and for very high SNRs, we observe a tendency for the highly perturbed training scenarios (CNR \nicefrac{1}{3}, 0) to be outperformed by the system trained with clean data (CNR $\infty$), which in turn is slightly outperformed by the slightly perturbed training scenarios (CNR 3, 1).

To assess whether these observations are statistically significant, we conducted a series of t-tests. 
This analysis was performed across four distinct CNR values. Table \ref{table:significant_differences} summarizes the ranges where statistically significant differences (\textit{t-statistic} < 0 and \textit{p-value} < 0.05) were observed in Precision, Recall, and F1 scores.

\begin{table}
\centering
\scriptsize
\begin{tabular*}{\columnwidth}{l|@{\extracolsep{\fill}}cccc@{}}

\toprule
\textbf{CNR} & \textbf{Precision (SNR)} & \textbf{Recall (SNR)} & \textbf{F1 Score (SNR)} \\ [.5mm]
\hline 
3, 1   & [0, 24]   & [-6, 15]  & [-6, 18]  \\ 
1/3    & [-3, 21]  & [-6, 15]  & [-6, 15]  \\ 
0      & [-6, 30]  & [-6, 9]   & [-6, 12]  \\ \bottomrule
\end{tabular*}
\caption{Statistically significant SNR ranges (\textit{t-statistic} < 0, \textit{p-value} < 0.05) in Precision, Recall, and F1 scores across different CNR levels.}
\label{table:significant_differences}
\end{table}

These findings indicate that while training with noise-injected data can enhance model robustness, excluding clean data entirely (CNR=0) leads to performance degradation in acoustically controlled environments, such as when testing on clean data (SNR=$\infty$). Thus, the absence of clean data during training adversely impacts performance under clean conditions. Moreover, the percentage of perturbed training samples is an important parameter of augmentation with impact the improvement gained and the SNR range of the improvement. 

\section{Conclusion}
This study highlights the importance of white noise injection in improving the robustness of APT models for diverse acoustic conditions. By introducing CNR, we show that under the right conditions, noise-injected training enhances performance at lower SNR levels and is on par at high SNRs. The findings emphasize the need for carefully selecting the data augmentation parameters and imply that increased robustness to different noise environments does not need to come at a cost for clean environments. In fact, there might be ways to gain significant improvements across all SNR levels.
Future work will investigate the impact on training data augmentation on other AMT systems, as well as extend data augmentation settings to be investigated. 

\bibliography{ISMIR2024_lbd}
\end{document}